\documentclass[
pra, 
aps,
twocolumn,
superscriptaddress,
longbibliography,
floatfix,
notitlepage]{revtex4-2}

\usepackage[utf8x]{inputenc}
\usepackage[english]{babel}
\usepackage[T1]{fontenc}
\usepackage{lmodern}

\usepackage{amsmath,amsfonts,amssymb,amsthm,bm,times,dcolumn,dsfont}

\usepackage{microtype}
\usepackage{braket}
\usepackage{gensymb} 
\usepackage{physics}

\usepackage[colorlinks={true}, citecolor={blue}, filecolor={blue}, linkcolor={blue}, urlcolor={blue}]{hyperref}
\usepackage{graphicx,color}
\usepackage[caption=false]{subfig}
\usepackage{tikz}
\usetikzlibrary{arrows.meta}
\begin{document}
\title{\bf{Quantum Coherent Route to Bath Induced Entanglement }}

\author{Kowsar. Al Mousavitaha}
\email{Kmousavitaha@gmail.com}
\affiliation{Department of Physics, Azarbaijan Shahid Madani
University, PO Box 51745-406, Tabriz, Iran}

\author{\"{O}zg\"{u}r E.~M\"{u}stecapl{\i}o\u{g}lu}
\email[Electronic address:]{omustecap@ku.edu.tr}
\affiliation{Ko\c{c} University, Department of Physics, Sar{\i}yer, Istanbul, 34450, T\"urkiye} 
\affiliation{TÜBİTAK Research Institute for Fundamental Sciences, 41470 Gebze, T\"urkiye}

\author{Esfandyar Faizi}
\email{efaizi.quantum@gmail.com}
\affiliation{Department of Physics, Azarbaijan Shahid Madani
University, PO Box 51745-406, Tabriz, Iran}

\date{\today}

\begin{abstract}
The micromaser is an archetype experimental setting where a beam of excited two-level atoms is injected into a high-finesse cavity. It has played a pivotal role as a testbed for predictions of quantum optics. We consider a generalized micromaser setting consisting of high-quality cavity pumped by a beam of three-level atoms. The atoms are assumed to be prepared to carry quantum coherence between their excited state doublet. Our objective is to produce quantum entanglement between the right-handed circular (RHC) and left-handed circular (LHC) polarized photons in the cavity, exploiting the quantum coherence in the pump atoms. For that aim, we derive the generalized micromaser master equation for our system. We find that the dynamics of the micromaser field driven by the pump beam is equivalent to two non-interacting RHC and LHC photonis systems sharing a common non-equilibrium environment. The effect of the shared bath is to mediate an incoherent interaction between the otherwise non-interacting cavity photons, which emerges only if the atoms carry quantum coherence. We take into account cavity losses as a source of quantum decoherence and characterize the quantum entanglement between the LHC and RHC polarized photons in terms of logarithmic negativity, calculated using the dynamical solution of the master equation. Our reseults reveal that while there is no steady-state entanglement, LHC and RHC polarzied photons can be entangled in the transient regime. 
\end{abstract}

\maketitle

\section{Introduction}\label{sec:intro}
Quantum entanglement lies at the heart of quantum information science and its applications in quantum computation~\cite{Nielsen,MLi}, simulation~\cite{TLiu}, communication~\cite{CPYang1, CPYang2, CPYang3, MHua}, metrology~\cite{Metrology} and sensing~\cite{Sensing}. While entanglement between two-level systems (qubits) has been routinely realized using available strong nonlinear interactions in various physical systems~\cite{GLCheng}, entanglement among larger systems, such as atomic Bose-Einstein condensates~\cite{Kumar}, optical cavity photons or vibrational phonons~\cite{YJZhao} of optomechanical resonators ~\cite{JLi}  turn out to be more challenging due their inherently linear, harmonic oscillator type character. 

The standard method to produce quantum entanglement between two optical resonator fields is to use photon hopping~\cite{MJHwang} by placing the cavities close to each other~\cite{Rehaily}. Another approach to entanglement generation is using an ancillary system mediating the coupling of two parties~\cite{MMariantoni, SLMa, RSharma}. The quantum decoherence challenges these coherent routes to entanglement~\cite{CPYang2, CPYang3} due to the physical environments in practical implementations. Noise-assisted preparation of entanglement ~\cite{XXYi} has been proposed  for a possible solution against the decoherence problem, which, however, requires precise control and engineering capabilities over environments, limiting its potential realizations~\cite{JJalali}. 
We propose an alternative strategy to produce quantum entanglement between non-interacting photonic modes. To this end, we consider a generalized micromaser system, where a beam of coherent three-level atoms passing through an optical cavity. We ask if the coherent atom beam can simulate a noisy but a beneficial environment that can entangle different polarization modes of the photons, even 
in the presence of cavity losses.

It has been shown that Gaussian states of a resonator field, such as thermal, coherent, or squeezed, can be generated by pumping a micromaser cavity by a beam of two-level atom clusters in quantum coherent superposition states~\cite{SMa}. The atomic beam simulates an effective environment for the resonator photons. The dynamics of the cavity photons can be described by a coarse-grained master equation, where the dissipative  terms correspond to effective environments that can be coherent, thermal, or squeezed-thermal depending on the type of quantum coherence in the atomic clusters. We generalize this idea to the case where a beam of three-level atoms drives a micromaser cavity (Illustrated in Fig~\ref{fig:setup}). The atoms interact with the cavity field in a V-type transition scheme. Depending on the atom's quantum state, we find that the action of the atomic beam on the cavities can play the role of a shared environment that can induce entanglement between the left-handed circular (LHC) and right-handed circular (RHC) polarized photons. The effective, shared environment compete against the harmful effects of cavity dissipation to the physical environment so that the resonator photons can evolve to an entangled state at the early stages of the dynamics. In the steady-state however the entanglement disapears due to the decoherence by the physical environment.

Our results can be translated to practical systems where timed interactions between bosonic and three-level subsystems can be realized. Such system specific extensions of our results can be significant for hybrid quantum information systems~\cite{PBLi, CPYang,YZhang, Rehaily} in superconducting resonators~\cite{FWStrauch} and solid state qubits~\cite{GLCheng}; quantum simulators in atomic BEC-cavity QED 
systems~\cite{Kumar}, and for studies of macroscopic quantum effects in optomechanical resonators.

The rest of the paper is organized as follows. In Sec.~\ref{sec:model}, we describe our generalized micromaser model and introduce the initial condition of the quantum coherent atoms forming the pump beam. We derive the master equation for the model system in Sec.~\ref{sec:masterEq} and present our interpreation how quantum coherence allows for bath mediated coupling between LHC and RHC polarzied photons. Our results of quantum entanglement dynamics are presented and discussed in~\ref{sec:results}. We conclude in Sec.~\ref{sec:conclusion}. 

\section{Micromaser pumped by quantum coherent three-level atoms}\label{sec:model}
\begin{figure}[h!]
\centering
\includegraphics[width=8.4cm, height=7.2cm]{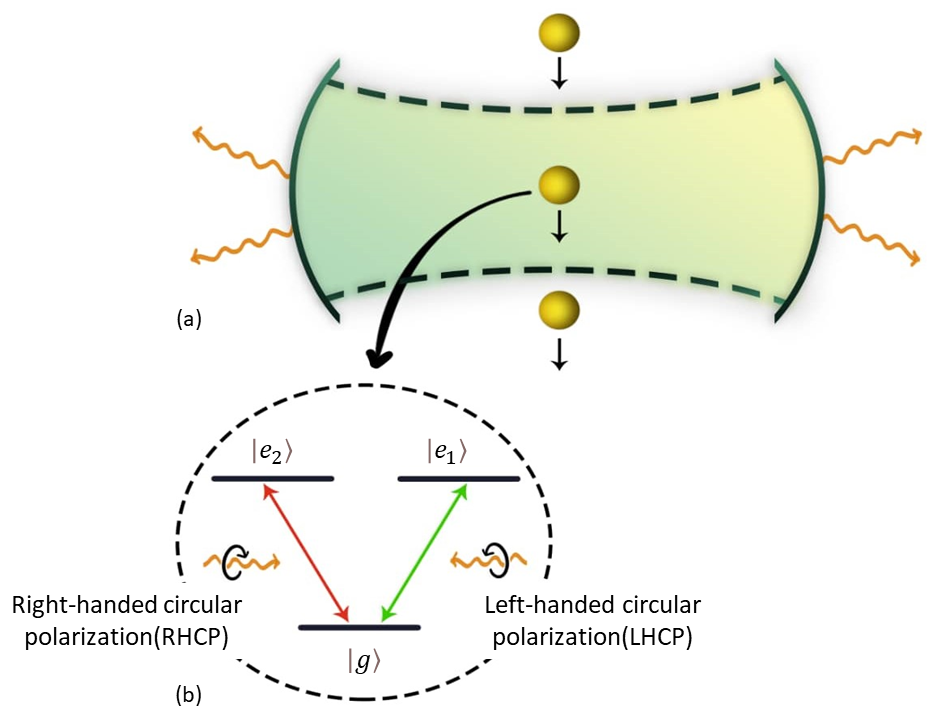}
\caption{(a) Typical micromaser setup, where an optical cavity is pumped by atoms. (b) We consider a simple generalization of usual two-level atomic pump used in typical  micromaser setups by considering a three-level atom.  The upper levels of the atom are assumed to be degenerate and denoted by states $\ket{e_{_1}}$, $\ket{e_{_2}}$, while the ground state is denoted by $\ket{g }$. We further assume a V-type transition scheme for the atom-cavity field interaction. The transitions  $\ket{e_{_1}} \leftrightarrow  \ket{g }$ and  $\ket{e_{_2}} \leftrightarrow \ket{g}$ are coupled to left-handed circular (LHC) and right-handed circular (RHC) polarized photonic modes of the cavity, respectively.}
\label{fig:setup}
\end{figure}

Let us consider a beam of V-type three-level atoms carrying quantum coherence between their upper states $\ket{e_{_1}}$ and $\ket{e_{_2}}$ passing through a micromaser cavity as depicted in Fig.~\ref{fig:setup}.

The atom-cavity field system is described by a Hamiltonian
\begin{eqnarray}
\hat H=\hat H_{A}+ \hat H_F+\hat H_I,
\end{eqnarray}
where the contributions of the free atom and the photons are written by (we take $\hbar=1$) 
\begin{align}
&\hat {H_A}= \sum_{i=1}^{2}\omega_0\ket{e_{i}} \bra {e_{i}},\\
&\hat {H_F}=\sum_{i=1}^{2}\omega_0{\hat a^\dag}_{i} \hat {a_{i}},
\end{align}
and
the the atom-cavity field (A-F) interaction in the dipole and rotating wave approximations is given by 
\begin{eqnarray}
\hat H_I  = \sum_{i=1}^{2}\hbar g_{i}(\hat a_{i} \ket{e_{i}}\bra{g}+{\hat a^\dag}_{i} \ket{g}\bra{e_{i}}).
\end{eqnarray}
The interaction Hamiltonian describes $V$-type transitions in the atom between the ground $\ket {g } $ and two excited states $\ket{e_{_1}}$ and $\ket{e_{_2}}$. We assume each transition is resonant with only one polarization mode
at the frequency $\omega_0$.  $\hat a_{i}, {\hat a^\dag}_{i}$ denote the annihilation and creation operators of photons in the LHC ($i=1$) and RHC ($i=2$) polarized mode. The coupling coefficients between the atom and the cavity modes $i=1,2$ denoted  by $g_{i}$. The model is  a multi-mode generalization of Jaynes-Commings model; it can be physically realizable in terms of angular momentum selection rules for the atomic Zeeman levels. 

From quantum thermodynamics perspective, our model generalizes the three-level atom with degenerate ground levels coupled to single-mode cavity to the case of two-mode cavity~\cite{Sun}. For single-mode case, the interest in Ref.~\cite{Sun} was to reveal how the quantum coherence in degenerate quantum levels can affect thermalization of the cavity field. Here, our objective is to explore if quantum coherence in the upper degenerate levels can be translated to produce quantum entanglement between the left- and right- polarization modes of the cavity.
A somewhat opposite question, how quantum coherence can be translated into heat to increase temperature of a photon gas inside a micromaser cavity has been examined in Ref.~\cite{Scully}, again when the quantum coherence is carried by the ground levels. In our case, instead of erasing quantum coherence or information to extract heat, our objective is to translate quantum coherence in atomic system to the
photonic system while promoting it to quantum entanglement. 

In a typical micromaser operation,  every atom in the pump beam is initially uncorrelated with the cavity photons, so that we can write the state of the total system just before the injection of an atom into the cavity region as a product state  
$\rho(0) = \rho _{_A} \otimes \rho _{_F}$ of the photonic  $\rho_{_F}$ and the atomic $\rho_{_A}$ states, where 
\begin{equation}
{ \rho _{{_A}}}(0) =  \begin{pmatrix}
{{p_{e_{_1}}}} & {\chi  \xi } & 0  \\ 
{ \chi  \xi } & {{p_{e_{_2}}}} & 0  \\ 
0 & 0 & {{p_g}}
\end{pmatrix}.
\end{equation}
Here, $p_{e_{_1}},p_{e_{_2}}$ and $p_g$ are the occupation numbers of the excited and  ground levels, respectively.  We assume there is quantum coherence between the upper levels, characterized by amplitude $\chi$.  For simplicity, we consider only real valued coherences. For a well-defined,
positive-definite, density matrix we have the conditions $\chi \le \sqrt {{p_{{e_{_1}}}}{p_{{e_{_2}}}}}$ and
${p_{e_{_1}}} + {p_{e_{_2}}} + {p_g} = 1$.
We introduce a
phenomelogical factor $0\le \xi\le 1$ to represent possible atomic dephasing of the quantum coherence. Due to the interactions with the earlier atoms, the photonic state cannot
be written as a product state of the correspoding cavity modes. Only before the arrival of the very first atom, we can write $\rho(0) = \rho _{_A}\otimes \rho _{_F}$, where $\rho_{_F}$ is the reduced density matrices of the cavity modes. 

\section{Open quantum system dynamics of polarization modes of the  micromaser}
\label{sec:masterEq}
This section closely follows the textbook derivation of micromaser master equation~\cite{Orszag}, generalized to our case the pump beam with the coherent three-level atoms. Let's first neglect the cavity losses, assuming that the atoms pass through the cavity region in a time $\tau$ much faster than cavity decay (photon loss) rates, so that a pump beam atom and cavity photons form a closed system. Assuming an atom, labeled with $j$, arrives at the cavity region at a time $t_j$, the initial state of the total atom-photons system $\rho(t_j)$ evolves unitarily to a state $\rho(\tau+t_j)$ when the atom leaves the interaction region according to 
$\rho (\tau+t_j) =  \hat U(\tau)\rho (t_j)\hat U^\dag(\tau)$. The propagator 
$ \hat U(\tau) = \exp \left( { - i{H_I }}t \right)$ is evaluated analytically in the interaction picture for given number of photons $m$ and $n$ in the cavity modes $1$ and $2$, respectively. It is expressed in the atomic energy basis $\{\ket{g },\ket{e_{_1}},\ket{e_{_2}}\}$ in the form
\begin{widetext}
\begin{equation}\label{eq:propagator}
U_{mn}(\tau) =  \begin{bmatrix}
\dfrac{{g_{_2}}^2n + {g_{_1}}^2m\cos \left( {\lambda \tau } \right)}{2{\lambda ^2}} &\dfrac { - {g_{_1}}{g_{_2}}\sqrt {mn} \sin \left( \frac{\lambda \tau } { 2} \right) } {\lambda ^2} &\dfrac 
{ - i{g_{_1}}\sqrt m \sin \left( {\lambda \tau } \right)}  {\sqrt 2 \lambda }  \\ 
\dfrac{ - {g_{_1}}{g_{_2}}\sqrt {mn} \sin \left(\frac{\lambda \tau }  {2} \right)} {\lambda ^2} & \dfrac{{g_{_1}}^2m + {g_{_2}}^2n\cos \left( {\lambda \tau } \right)}  {2{\lambda ^2}} &\dfrac { - i{g_{_2}}\sqrt n \sin \left( {\lambda \tau } \right)}  {\sqrt 2 \lambda }  \\ \dfrac
{ - i{g_{_1}}\sqrt m \sin \left( {\lambda \tau } \right)}  {\sqrt 2 \lambda } &  \dfrac{ - i{g_{_2}}\sqrt n \sin \left( {\lambda \tau } \right)}  {\sqrt 2 \lambda } & {\cos \left( {\lambda \tau } \right)}
\end{bmatrix},
\end{equation}
\end{widetext}
where, we introduce $\lambda=[(g_{_1}^2m + g_{_2}^2n) / 2]^{1/2} $. 

The density matrix of the cavity photons $\rho _{F}$  after the interaction with the atom can be written as
\begin{eqnarray}
\rho _{_F}( t_j + \tau ) = M( \tau )\rho_{_F} (t_j),
\end{eqnarray} 
where the map $M(\tau)$ is defined as a superoperator 
\begin{equation}\label{eq:map}
M(\tau )\rho _{_F}(t_j) = \Tr_{A}[ \hat U( \tau)\rho_{_A}(t_j)
\otimes \rho_{_F}(t_j) \hat U^\dag (\tau ) ].
\end{equation}
Here, $\Tr_{_A}$ means tracing over the atomic degrees of freedom.

Let's assume that the atoms arrive at the cavity region at random times; in other words, the series of atom-photon interactions are discrete events that we do not know when they happen precisely. However, we suppose that we know the mean time between the interactions so that we can introduce a collision or interaction rate $r$.
As the arrivals of atoms to the cavity region are uncorrelated, the sequential interactions in our case are classified as a Poisson process, characterized by a rate $r$. Hence,  the probabilities for an atom to be or not to be in the interaction region in a time interval $(t+\Delta t)$ are given by $r\Delta t$ and $1-r\Delta t$, respectively. Only in the former case, the state of the photons will be updated according to the map in Eq.~\ref{eq:map}, while in the latter case, the state remains the same, hence we write
\begin{eqnarray}
\rho_{_F}(t+\Delta t)=r\Delta t M(\tau )\rho _{_F}(t) + (1-r\Delta t)\rho _{_F}(t).
\end{eqnarray}
For $\Delta t$ shorter than any time scale of interest, we can express the dynamics of the photons in the same form of standard coarse-grained 
micromaser master equation
\begin{eqnarray}\label{eq:master}
\dot{\rho_{_F}}=r[ M( \tau ) - 1]\rho_{_F}.
\end{eqnarray}
Using Eq.~(\ref{eq:master}) and expanding the propagator in Eq.~(\ref{eq:propagator}) to second order in $\lambda\tau$ for sufficiently small interaction times, and neglecting the cavity loss in the short interaction time, we obtain the coarse-grained micromaser master equation for our
system to be
\begin{widetext}
\begin{eqnarray}\label{eq:Master1}
\dot{\rho}_{_F}
=\gamma_{_1}p_gD[ \hat a_{_1} ]\rho _{_F} +\gamma_{_2}p_gD[ \hat a_{_2} ]\rho _{_F}+ \gamma_{_1}p_{e_{_1}}D[ {\hat a^\dag} _{_1}]\rho _{_F} + \gamma_{_2}p_{e_{_2}}D[ {\hat a^\dag}_{_2} ]\rho _{_F}
+\gamma_{_{12}}\chi\xi  \big(D[  {\hat a^\dag}_{_1} ,{\hat a^\dag}_{_2}]+
D [  {\hat a^\dag}_{_2},{\hat a^\dag}_{_1} ]\big)\rho _{_F},
 \end{eqnarray}
\end{widetext}
where the dissipator superoperators are given by
\begin{eqnarray}
D[\hat x,\hat y ]\rho&=&\Big(\hat x \rho \hat y^\dag - \frac{1}{2}
\{ \hat x^\dag \hat y ,\rho\} \Big),\\
D[ \hat x] \rho&=&\Big(\hat x \rho \hat x^\dag - \frac{1}{2}
\{ \hat x^\dag \hat x ,\rho\} \Big).
\end{eqnarray}
We denote the effective bath coupling rates by 
\begin{eqnarray}
\gamma_{_1}=\frac{1}{2}r\tau^2{g_{_1}}^2,\gamma_{_2}=\frac{1}{2}r\tau^2{g_{_2}}^2, \gamma_{_{12}}=\frac{1}{2}r\tau^2{g_{_1}}{g_{_2}}. 
\end{eqnarray}

The master equation can be interpreted as an open quantum system dynamics of two distinct photon gases in a shared environment. The pump atoms act as an effective engineered shared environment. All the dissipation channels of the environment are controllable by the populations and the coherence in the pump atoms. The first four terms in the master equation~(\ref{eq:Master1}) represent the
effective local bath action of the atomic beam on the photon gases. The local dissipation channels are controlled by the initial population of the ground and upper levels of the three-level atoms in the driving beam. The first two terms are for the energy absorption processes of the driving atoms from the resonator fields. When the ground state is dominantly populated, the effect is analogous to the decay of the cavity fields into a low or zero temperature environment. The effective environment-induced excitations of cavity fields are described in the third and the fourth terms. They are controlled by the initial populations of the excited states. 
The last two terms are dramatically different from these four local noise channels.   They describe a bath-mediated (incoherent) interaction between otherwise 
uncoupled cavity polarization modes by a shared effective bath simulated by the sequential drive of the cavity by the three-level atoms. In contrast to local noise channels, these latter terms represent not the harmful but the beneficial effects of the effective environment. This global effect of the effective environment only arises when the initial three-level atom has quantum coherent upper levels. 

Our objective is to determine if the effective bath-mediated interaction by the driving atoms can induce sufficiently strong quantum correlations between the RHC and LHC polarization modes of the resonator field to put them in an entangled state, even in the presence of local noise and additional decoherence channels. In addition to the effective bath local noise terms, the cavity is an open systems in its natural environment, and hence we should take into account additional terms to our master equation. Denoting the loss rates for LHC and RHC polarized photons by $\kappa_{_1}$ and $\kappa_{_2}$, respectively, Eq.~\ref{eq:Master1} changes to
\begin{equation}\label{eq:Master2}
\dot{\rho}_{_F}( t )
=r[ {M( \tau  )-1} ]\rho _{_F}(t)+\kappa_{_1}D[ \hat a_{_1} ]\rho _{_F} +\kappa_{_2}D[ \hat a_{_2} ]
\rho _{_F}.
\end{equation}

\section{Dynamics of Quantum Entanglement}
\label{sec:results}

In this section, we shall discuss the quantum dynamics of the entanglement between the polarization modes of the cavity. The two modes are assumed to be initially in the Fock state $\ket {{\psi _0}} = \ket{ 1 }_{a_{_1}}\otimes\ket{ 0}_{a_{_2}} $. 
This is not an arbitrary choice but such an initial condition, where one of the cavity modes is excited while the other one is empty, allows for efficient exchange of excitations between the polarization modes mediated by the atomic beam so that optimal quantum entanglement can build between the polarization modes in early time.
An initial external drive on the cavity can be envisioned in practice to prepare such a polarized single-photon state of the cavity field.
 
\begin{figure}[t!]
	
	\subfloat[\label{f0a}]{%
		\includegraphics[width=\columnwidth]{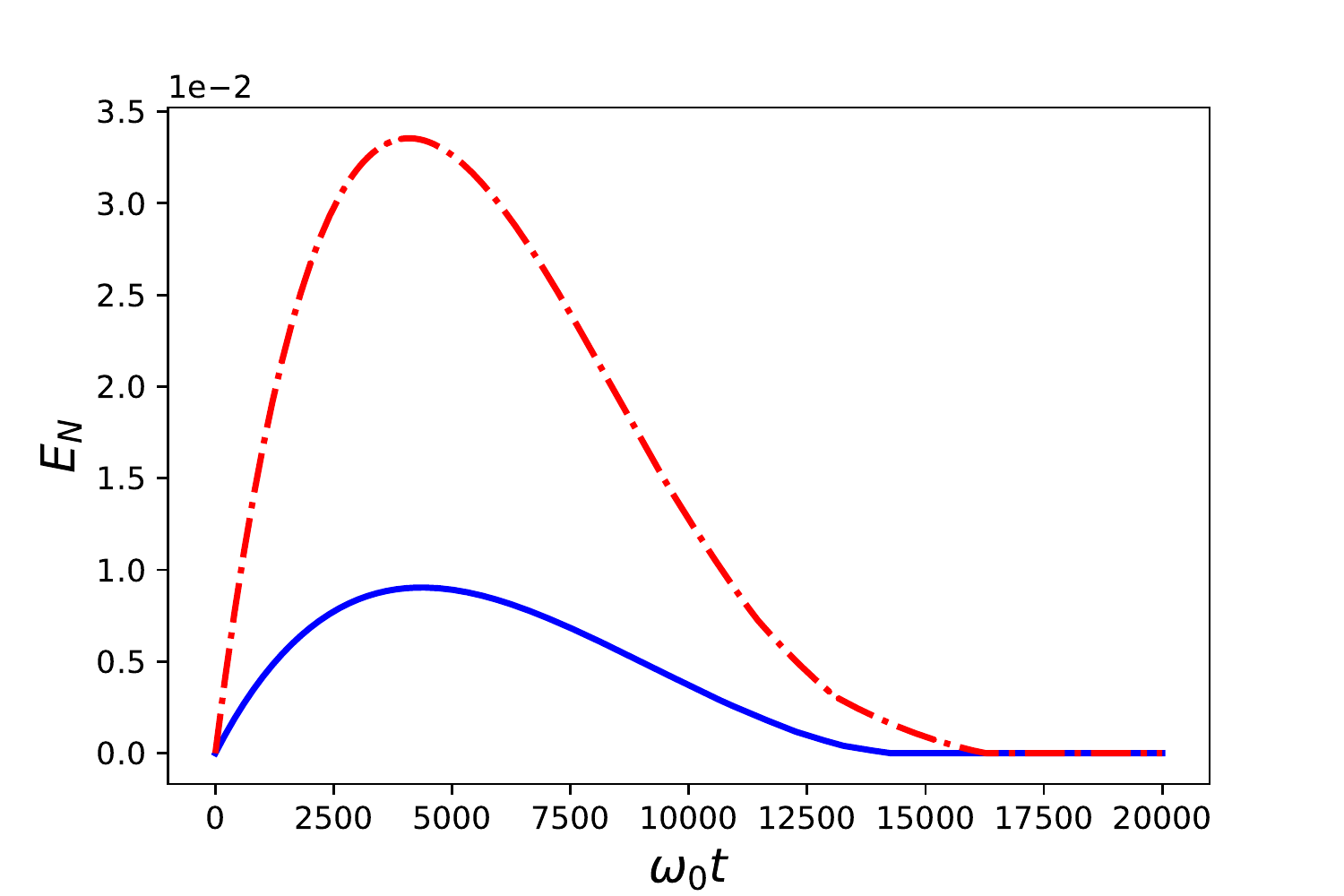}%
	}
	\\
	\subfloat[\label{f14b}]{%
		\includegraphics[width=\columnwidth]{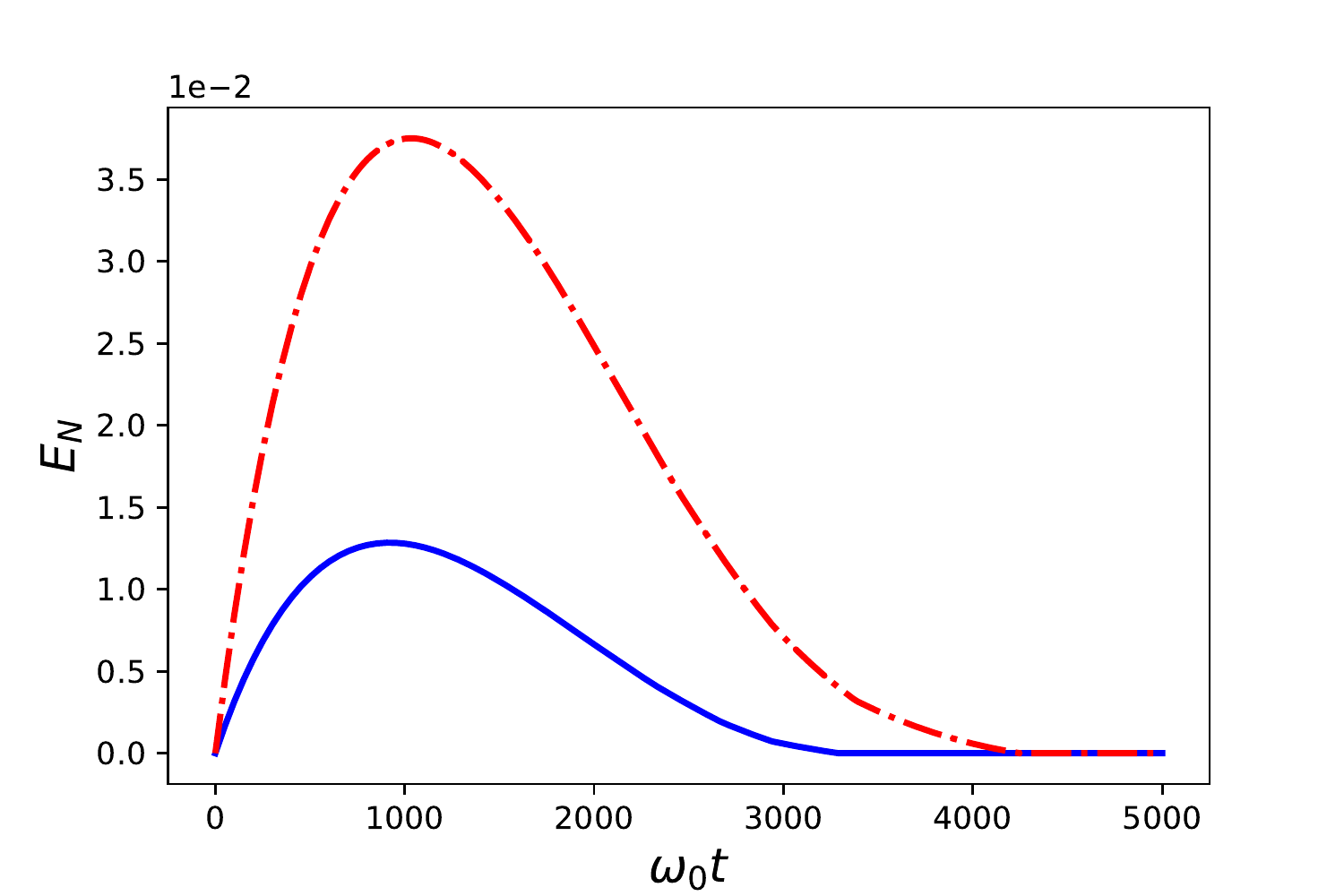}   %
	}
		\caption{Time dependence of the logarithmic negativity $E_N$ of resonator modes initially in a Fock state for $g_{_1}=9\times10^{-2} \omega_0, g_{_2}=5\times10^{-2} \omega_0$, $\kappa_{_1}=10^{-6} \omega_0$  and $\kappa_{_2}=2\times10^{-6} \omega_0 $, $p_{e_{_1}}={5}/{8}$, $p_{e_{_2}}={5}/{16}$, $p_{g}={1}/{16}$. The blue solid lines correspond to $\xi=0.7 $ and red dashed-dotted lines for $\xi=0.8 $, for (a) $r=0.1 \omega_0$ and (b) $r=0.5 \omega_0$. These parameters scaled with $\omega_0=10^{10}\text{Hz}$.}
 
		\label{fig-2}
	
\end{figure}

\begin{figure}[t!]
	
	\subfloat[\label{f14a}]{%
		\includegraphics[width=\columnwidth]{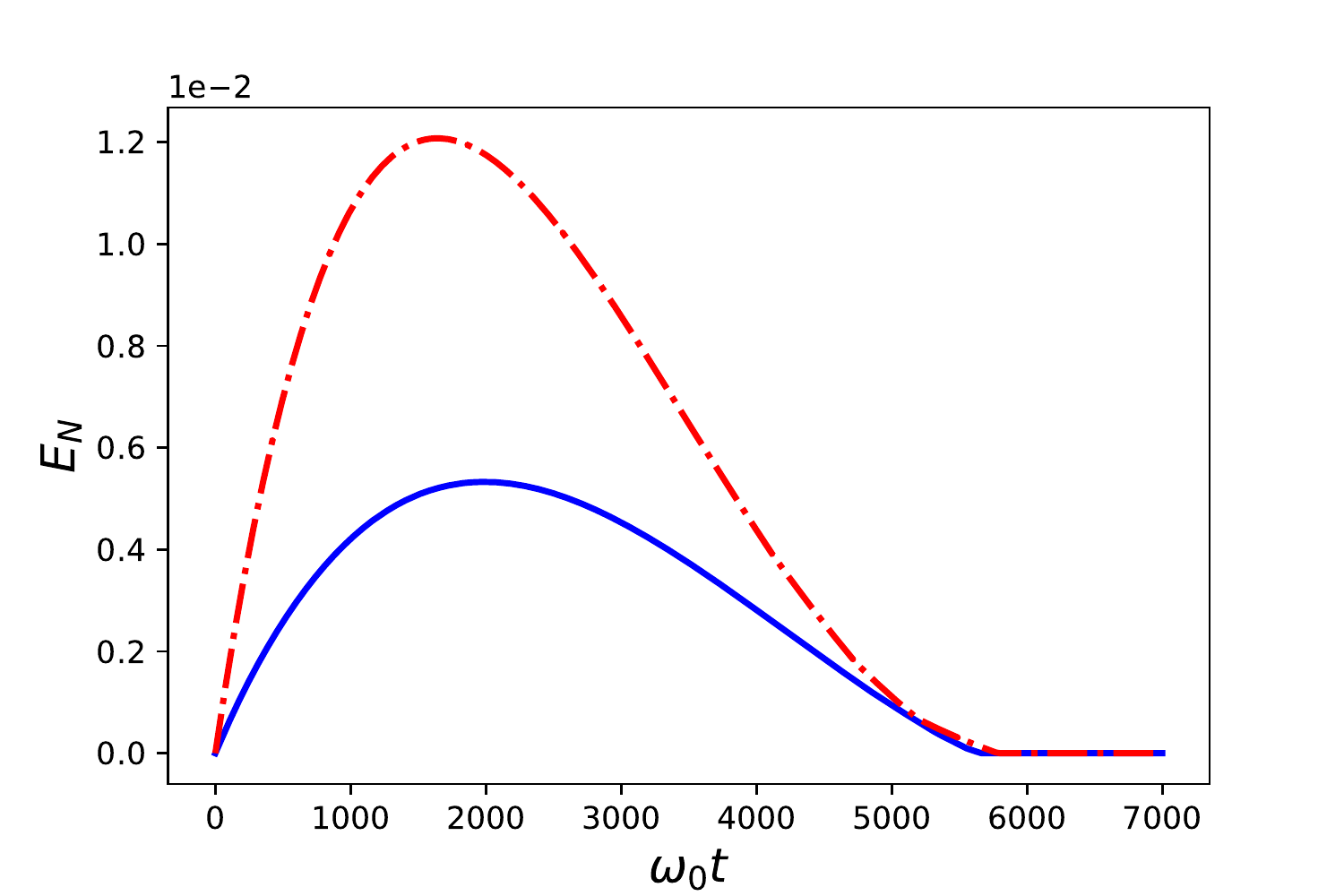} %
	}
	\\
	\subfloat[\label{f14b}]{%
		\includegraphics[width=\columnwidth]{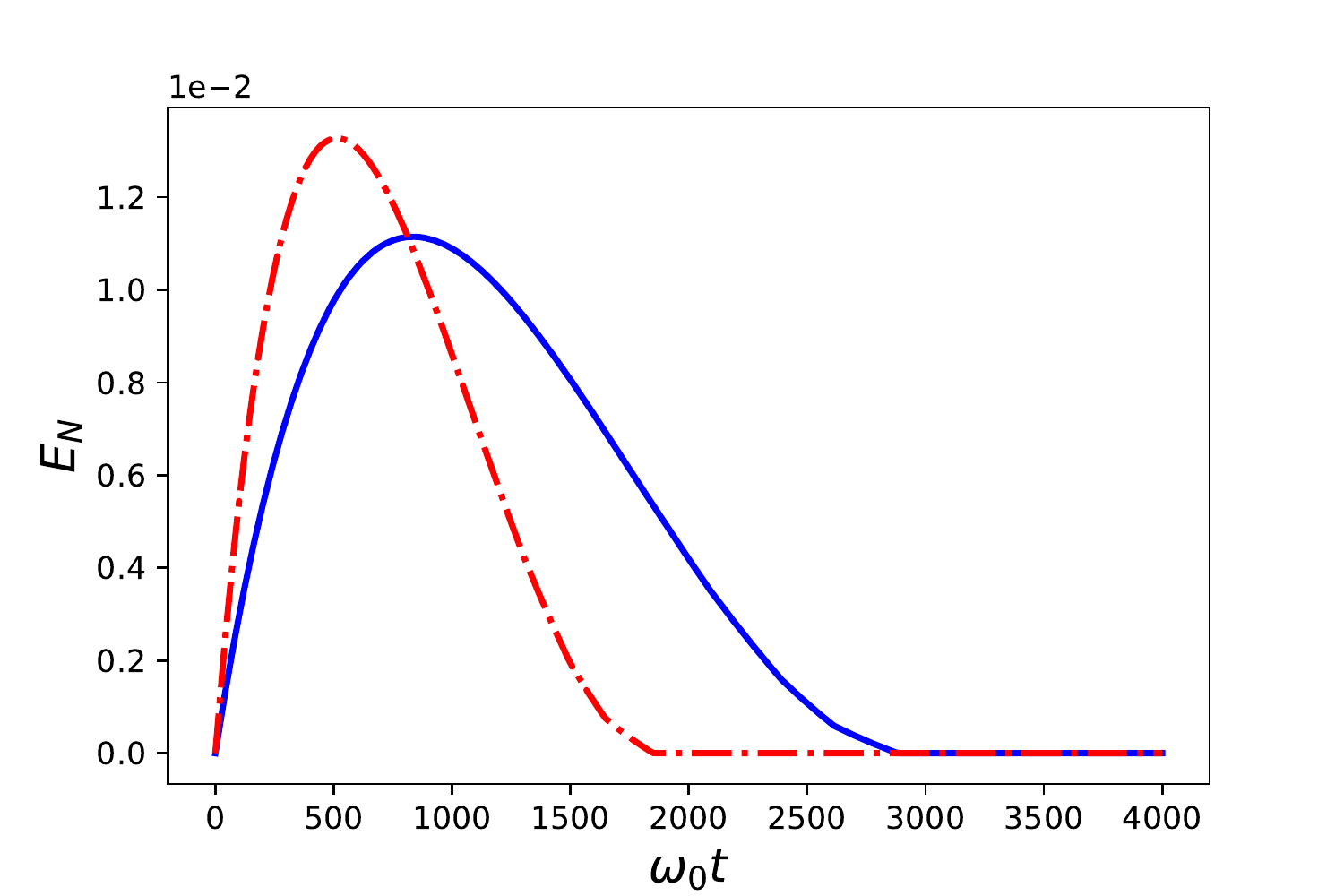}   %
	}
	\caption{Time dependence of $E_N$  for $\xi=0.7$. The blue solid lines correspond to $,  r=0.1 \omega_0$  and red dashed-dotted lines for $r=0.5 \omega_0$. For (a) $g_{_1}=g_{_2}=5\times10^{-2}\omega_0$ and, (b) $g_{_1}=g_{_2}=9\times10^{-2} \omega_0$. Other parameters are the same as Fig.~\ref {fig-2}. These parameters scaled with $\omega_0$.}
	\label{fig-3}

\end{figure}

We examine the entanglement between the two polarization mode of the cavity using the the logarithmic negativity $E_N\left(\rho  \right)$~\cite{{Vidal}}. The logarithmic negativity is a computable and a nonconvex entanglement monotone, which is given by 
\begin{equation}
{E_N}\left( \rho  \right): = {\log _2}\left\| {{\rho ^{T_{A}}}} \right\|.
\label{eq:negativity}
\end{equation}
Here, $\rho ^{T_{A}}$ denotes the partial transpose of $\rho$ with respect to party A and $\left\| {{\rho ^{T_{A}}}} \right\|$ is the trace norm of $\rho ^{T_{A}}$.

To compute the logarithmic negativity in Eq.~(\ref{eq:negativity}), we evolve the initial state numerically by solving the master equation Eq.~\ref{eq:Master2} using QuTip 4.6 ~\cite{QuTiP}. Our numerical calculations are performed by using the typical
parameter values for superconducting circuit quantum electrodynamics systems (QED). We emphasize that
our generic approach is applicable to any optomechanical, microwave, or optical resonators. We consider the case of superconducting resonators specifically as they are most promising to get significant quantum entanglement due to their strong coupling and low decoherence conditions~\cite{Vidal, Plenio, Sun}.
Accordingly, we take $g_{_1} = 9  \times 10^{8} \text{Hz}$, $g_{_2} = 5  \times 10^{8} \text{Hz}$, $r=  10^{9} \text{Hz}$, $\kappa_{_1} = 1\times 10^4 \text{Hz} $, $\kappa_{_2} = 2\times 10^4 \text{Hz}$ and  $\tau=10^{-10}s$. 
We use the atomic and cavity resonance frequency $\omega_0$ as our scaling parameter such that all the physical quantities become dimensionless and scaled in the unit system   $\hbar \omega_0=1, \hbar=1, \omega_0=1$.

Our results are presented in Fig.~\ref{fig-2} and ~\ref{fig-3}  which plots the 
time evolution of the $E_{N}$ for different values of $\xi,r,g_{_1}$ and $g_{_2}$. The general conclusion is that the quantum coherence injected into the cavity field by the atomic beam is not sufficient to refine into quantum entanglement between the polarization modes in the steady state. However, we can get significant quantum entanglement in the transient regime of the open system dynamics. 

Fig.~\ref{fig-2} specifically indicate that entanglement is increasing with the atomic coherence ; while Fig.~\ref{fig-3} also shows that entanglement is
increasing with the atom-cavity coupling strength. Hence in designing the repeated interactions, one should prepare conditions to minimize the atomic dephasing ($\xi\rightarrow 1$ for our case) and increase the atom-cavity interactions. In micromaser-type repeated interaction settings, the latter condition can be achieved by increasing the arrival rate of the atoms to the cavity region as shown in 
Fig.~\ref{fig-2} and ~\ref{fig-3} . 

In our present approach we take the maximum coherence in
the upper doublet of  three-level atoms and characterize the effect of smaller values using the atomic dephasing parameter. An appealing possibility to increase the quantum coherence could be to use atomic clusters instead of single atom in the pump beam. This could allow for superradiance type of quadratic scaling in the strength of the effective bath mediated coupling of the polarization modes. We can expect
faster and stronger entanglement of the cavity photons as a result of cooperative
effort fo atoms in the clusters. 

Fig.~\ref{fig-2} presents the case of non-equal cavity-atom coupling strengths. 
The bath mediated coupling is symmetric in $g_{_1}$ and $g_{_2}$ exchange, and
proportional to the product $\sim g_{_1} g_{_2}$. Accordingly, 
the entanglements reported in Figs.~\ref{fig-2} is consistent
with the hierarchy $g_{_1} > g_{_2}$ so, $E_N$ in Fig.~\ref{fig-2} lies in between
$E_N$ of Figs.~\ref{fig-3} (a) and (b).

\section{Conclusion }
\label{sec:conclusion}
We addressed the question of how to entangle two polarization modes of a micromaser cavity field. Our solution is based upon a generalization of the micromaser setting where a beam of atoms pumps an optical resonator. Our conception uses a pump beam of three-level atoms with quantum coherence in degenerate upper-level doublet. By deriving the micromaser master equation for such a pump-beam, we point out that the atomic beam acts as an effective, shared bath for cavity modes, and the bath-mediated coupling of the polarization modes is proportional to the atomic coherence. However, the atomic dephasing and cavity dissipation challenge the translation of atomic coherence to photonic entanglement. We tested our idea using the typical parameter values of superconducting transmission line resonators. The entanglement emerges in the transient regime but disappears in the steady-state. A possible improvement could be considering atomic clusters instead of a single three-level atom for scaling of coherence induced bath mediated coupling of polarization modes. We hope our proposal can inspire studies for scalable multi-mode photonic entanglement sources for diverse quantum technology applications.

\section*{Acknowledgment }
The authors gratefully acknowledge Seyed Mahmoud Ashrafi, Muhammad Tahir Naseem, and Mohsen Izadyari  for fruitful discussions. \"{O}~.E.~M. acknowledges  support from TUBITAK Grant No.~120F230.
\bibliographystyle{apsrev4-2}
\bibliography{refpaper10} 
\end{document}